\documentclass[11pt]{article}

\def\tr{\;{\rm tr}\;}

\def\bra{\langle}   \def\ket{\rangle}

\def\implies{\Rightarrow}

\newcommand{\tl}[1]{\tilde{#1}}

\newcommand{\dd}[2]{\frac {\partial #1}{\partial #2}}
\newcommand{\pdr}{\partial}

\newcommand{\beq}{\begin{eqnarray}}
\newcommand{\eeq}{\end{eqnarray}}

\newcommand{\half}{\frac{1}{2}}

\newcommand{\ov}[1]{\frac{1}{#1}}

\def\a{\alpha}      \def\gb{\beta}   \def\g{\gamma}       
\def\gd{\delta}      \def\D{\Delta}  \def\eps{\epsilon} 
                \def\La{\Lambda}

        \def\tht{\theta}  
              \def\ch{\chi}
\def\om{\omega}

\newcommand{\Ntr}{{{\rm tr} \over N}}

\newtheorem{sutra}{}

\newtheorem{bhashya}{}[sutra]

\def\spec{{\rm spec}}

\input{epsf}
\usepackage{hyperref}

\begin{document}


\begin{titlepage}

\title{\Large\bf
Schwinger-Dyson operators as invariant vector fields on a matrix-model analogue of the group of loops}

\author{Govind S. Krishnaswami}
\date{\normalsize
Department of Mathematical Sciences \& Centre for Particle Theory, \\
Durham University, Science Site, South Road, Durham, DH1 3LE, UK \vspace{.2in}\\
Chennai Mathematical Institute, \\
Padur PO, Siruseri 603103, India.
\smallskip \\ e-mail: \tt govind.krishnaswami@durham.ac.uk \\ 4 March, 2008}

\maketitle

\begin{quotation} \noindent {\large\bf Abstract } \medskip \\

For a class of large-$N$ multi-matrix models, we identify a group $\bf G$ that plays the same role as the group of loops on space-time does for Yang-Mills theory. $\bf G$ is the spectrum of a commutative shuffle-deconcatenation Hopf algebra that we associate to correlations. $\bf G$ is the exponential of the free Lie algebra. The generating series of correlations is a function on $\bf G$ and satisfies quadratic equations in convolution. These factorized Schwinger-Dyson or loop equations involve a collection of Schwinger-Dyson operators, which are shown to be right-invariant vector fields on $\bf G$, one for each linearly independent primitive of the Hopf algebra. A large class of formal matrix models satisfying these properties are identified, including as special cases, the zero momentum limits of the Gaussian, Chern-Simons and Yang-Mills field theories. Moreover, the Schwinger-Dyson operators of the continuum Yang-Mills action are shown to be right-invariant derivations of the shuffle-deconcatenation Hopf algebra generated by sources labeled by position and polarization.

\end{quotation}

\vfill \flushleft




\thispagestyle{empty}

\end{titlepage}

\eject

{\footnotesize \tableofcontents}

\section{Introduction}
\label{s-intro}

Quantum Yang-Mills theory is at the heart of the microscopic description of strongly interacting particles. The group of based loops on space-time plays an important role in the formulation of Yang-Mills theory in terms of Wilson loops, which are gauge invariant variables containing much physical information\cite{gambini-pullin-book}. Expectation values of Wilson loop observables are functions on this group. Expansion around the multi-color limit is a promising approximation method to solve Yang-Mills theory \cite{thooft-large-N}. However, in the absence of a full-fledged differential geometry and analysis on the space of loops, progress in understanding and approximately solving the multi-color limit of Yang-Mills theory has been partly held up despite important early work \cite{thooft-large-N,witten-baryons-N,migdal-phys-rpts}.

On the other hand, hermitian multi-matrix models may be regarded as toy-models for Yang-Mills theory. The $N \times N$ matrices may be thought of as gauge fields at various space-time points, where $N$ is the number of colors. It is natural to ask whether there is an analogue of the group of loops such that multi-matrix correlators are functions on this group. If so, can we interpret the factorized Schwinger-Dyson loop equations (fSDE)\footnote{The fSDE are quantum-corrected equations of motion for a matrix model in the multi-color (large-$N$) limit. They are analogous to the Makeenko-Migdal equations\cite{migdal-phys-rpts} of large-$N$ Yang-Mills theory. $N=3$ in nature.} of a matrix model in terms of the differential operators and products naturally associated to this group? Doing so may open up new perspectives and approximation methods for the large-$N$ limit of multi-matrix models and Yang-Mills theory.

In this paper we show that there is indeed such a group $\bf G$ associated to a large class of multi-matrix models. We construct it indirectly as the group of characters of the commutative shuffle-deconcatenation Hopf algebra\footnote{Technically, ${\bf G}$ can be thought of as an analogue of the group of generalized loops or extended loop group studied by Tavares\cite{tavares} and Bartolo, Gambini and Griego \cite{bartolo-gambini-griego}. We also identify a subgroup of ${\bf G}$ which can be regarded as an analogue of the smaller group of loops on space-time, see sec. \ref{s-mat-mod-analogue-of-loop-gp}.}. The latter is defined using the shuffle and concatenation products and reversal of order of matrices in a correlator. These are analogous to point-wise products of functions of loops, concatenation of loops, and reversal of loop orientation. In the simplest case of a single matrix and real-valued characters, ${\bf G}$ is the multiplicative group of non-zero real numbers. More generally, $\bf G$ is identified with the exponential of the free Lie algebra. We develop the rudiments of differential calculus on ${\bf G}$ using algebraic operations in the Hopf algebra. We find a large class of (formal) matrix models that can be formulated in terms of this group. We show that their fSDE are quadratic equations (in the convolution product) for a function on $\bf G$. Moreover, there is one equation for each linearly independent primitive element of the Hopf algebra of functions on $\bf G$. The Schwinger-Dyson (SD) operators, one for each linearly independent primitive, are shown to be right-invariant vector fields on $\bf G$. Thus, given a prescription of which right-invariant vector field to associate to a given primitive, we can write down a system of fSDE for any group. For the group of relevance to matrix models, this prescription is encoded in the action. We find a large class of admissible actions and their SD operators, which include the gaussian, Chern-Simons and Yang-Mills matrix models as special cases. Finally, the SD operators of the Yang-Mills action are obtained and shown to be right-invariant derivations of the shuffle-deconcatenation Hopf algebra on a continuously infinite number of generators labeled by space-time position and polarization. However, in the case of Yang-Mills theory, we still need to pass to a quotient of this Hopf algebra to account for gauge invariance and recover the group of loops\cite{chen-bulletin,tavares} (or need to gauge-fix and introduce additional generators for ghosts\cite{ghost-gluon}), before we can look for physical solutions to the fSDE.

In \cite{entropy-var-ppl} the fSDE were formulated as conditions for the extremum of a large-$N$ `classical' action (the Legendre transform of the entropy of operator-valued random variables). This viewpoint applied to generic multi-matrix models and also provided a variational approximation method. Here we develop a quite different group theoretic formulation, which only applies to a sub-class of matrix models. The distinction is very roughly analogous to that between the generic classical mechanical system and one whose configuration space is a group. However, due to this restriction, we find additional structures which closely mimic those present in Yang-Mills theory. We wanted to study these structures since they form the basis for an approximation scheme for multi-matrix models proposed in \cite{deform-prod-der}. We hope our group theoretic formulation allows for a generalization to more familiar groups, where solutions to the fSDE may be more easily found. Our work continues the developments in the physics literature due to Migdal and Makeenko \cite{migdal-phys-rpts,makeenko-book}, Polyakov \cite{Polyakov}, Cvitanovic et. al. \cite{cvitanovic-et-al}, Gambini et. al. \cite{gambini-pullin-book,bartolo-gambini-griego}, Tavares \cite{tavares}, Rajeev et. al. \cite{rajeev-turgut-der-free-alg,entropy-var-ppl}, and builds on our previous papers \cite{deform-prod-der,ghost-gluon}. There are of course other approaches to multi-matrix models such as those related to integrable models and algebraic geometry, see for instance \cite{kazakov-marshakov,eynard-loop-eqn-chain}.

By a $\La$-matrix model \cite{bipz,entropy-var-ppl,deform-prod-der,non-anom-ward-id} we mean a statistical system whose variables are a collection of random hermitian $N \times N$ matrices $A_i$, $1 \leq i \leq \La$ with partition function $Z = \int dA e^{- N \tr S(A)}$. The integration is over all independent matrix elements. The action is the trace of a polynomial in the matrices $\tr S(A) = \tr S^I A_I$ where\footnote{Capital letters denote multi-indices and repeated indices are summed. If $I= i_1 i_2 \cdots i_n$ then $S^I = S^{i_1 \cdots i_n}$ and $A_I = A_{i_1} A_{i_2} \cdots A_{i_n}$ is a product of $n$ matrices. The Kronecker $\gd^I_J$ is equal to one if $I=J$ and zero otherwise.} $S^I$ are the coupling `tensors' (see sec. \ref{s-fSDE-on-grp-spec-La} for examples). The $A_i$ model gauge fields at a collection of space-time points labeled by $i$ and are $N \times N$ matrices in color space. Observables $\Phi_I = \Ntr A_{I}$ are invariant under the global $U(N)$ action $A_i \to U A_i U^\dag$. We are interested in their expectation values in the large-$N$ limit, the cyclic tensors ($\emptyset$ denotes the empty word)
    \beq
    G_I = \bra \Ntr A_I \ket = \lim_{N \to \infty} \ov{Z} \int dA e^{-N \tr S(A)} \Ntr A_I; ~~~~~ G_\emptyset = 1.
    \eeq
$G_I$ satisfy a closed system of factorized\footnote{Factorization\cite{makeenko-book} is the property $\bra \Phi_{I_1} \Phi_{I_2} \cdots \Phi_{I_n} \ket = G_{I_1} \cdots G_{I_n} + {\cal O}(\ov{N^2})$ as $N \to \infty$.} Schwinger-Dyson equations, conditions for the invariance of $Z$ under infinitesimal non-linear changes of integration variable. The latter are infinitesimal automorphisms of the tensor algebra generated by $A_i$, $L_I^i A_j = \gd_j^i A_I$. The fSDE one for each letter $i$ and word $I$, relate a change of action to a change in measure
    \beq
    S^{J_1 i J_2} G_{J_1 I J_2} = \gd_I^{I_1 i I_2} G_{I_1} G_{I_2}.
    \eeq
If we define $\La$ non-commuting sources $\xi^i$, we can form the generating series of correlators $G(\xi) = \sum_I G_I \xi^I$. Then\footnote{Juxtaposition $\xi^{I_1} \xi^i \xi^{I_2}$ denotes concatenation $\xi^{I_1 i I_2}$.} the fSDE can be written ${\cal S}^i G(\xi) = G(\xi) \xi^i G(\xi)$. The SD operators
    \beq
    {\cal S}^i = \sum_{n \geq 0} (n+1) S^{i j_1 \cdots j_n} D_{j_n} \cdots D_{j_1}
    \eeq
are expressed in terms of left annihilation operators $D_j$ which satisfy $D_j \xi^{i_1 \cdots i_n} = \gd_j^{i_1} \xi^{i_2 \cdots i_n}$ or equivalently, $[D_j G]_I = G_{jI}$. For more details on the fSDE, we refer to \cite{entropy-var-ppl,deform-prod-der,non-anom-ward-id}. In this paper we do not have anything to say about the convergence of matrix integrals. We only use them as a formal device to generate the fSDE, whose structure we wish to investigate.

\section{Hopf algebra structure on correlations}
\label{s-hopf-on-corrlns}

The space of based oriented loops ${\gamma}$ on space-time (up to equivalence under backtracking or retracing), plays a basic role in the Wilson loop formulation of Yang-Mills theory. This loop space forms an infinite dimensional non-abelian group, with successive traversal of loops as product $\g_1 \g_2$ and reversal of orientation $\bar \g$ as inverse. The information in this group of loops can be encoded in the algebra of (complex-valued) functions defined on it. Wilson loop functions $W(\gamma) = \tr P\exp \oint A_\mu(\gamma(t)) \dot \gamma^\mu(t) dt$ (trace of holonomy of the gauge connection $A_\mu(x)$ around the closed loop $\gamma^\mu(t)$) form an adequate class of functions for this purpose \cite{gambini-pullin-book}. Since the underlying loop space is a non-abelian group, the algebra of functions has the additional structure of a commutative but non-cocommutative Hopf algebra (under suitable hypotheses, this is a general property of the algebra of functions on any group \cite{cartier-primer}). The point-wise product is $(W_1 W_2)(\g) = W_1(\g) W_2(\g)$, the coproduct $(\D W)(\g_1, \g_2) = W(\g_1 \g_2)$ encodes the concatenation of loops and the antipode $(SW)(\gamma) = W(\bar \g)$ encodes the inverse. The product and coproduct define compatible algebra and coalgebra structures, while the antipode turns this bialgebra into a Hopf algebra. Up to some technicalities, the underlying group of loops can be recovered as the spectrum (group of characters) of this Hopf algebra \cite{cartier-primer,abe-hopf-algebras,tavares}.

For a multi-matrix model we did not know the analogue of the group of loops, but we did notice a bialgebra structure on the multi-matrix correlators $G_I$ in connection with some approximation schemes for the loop equations \cite{deform-prod-der}. We recall this bialgebra structure and then define a compatible antipode to obtain the shuffle-deconcatenation Hopf algebra, which is the analogue of the Hopf algebra of Wilson loop functions. In the next section we will extract the underlying group from this Hopf algebra. The shuffle-deconcatenation Hopf algebra has appeared previously in other contexts \cite{reutenauer,cartier-primer,tavares}.

Let $G(\xi) = \sum_I G_I \xi^I$ denote the generating series of multi-matrix correlators in the large-$N$ limit. $G(\xi)$ is an element of the vector space ${\bf C}\bra \bra A \ket \ket$ of formal complex linear combinations of words $\xi^I$ in generators chosen from the alphabet $A = \{\xi^{i}, 1 \leq i \leq \La \}$ consisting of sources $\xi^i$, one for each matrix $A_i$. The commutative shuffle product $sh$ of two such series is $(F \circ G)(\xi) = \sum_{I} (F \circ G)_I \xi^I$ where $(F \circ G)_I = \sum_{I = J \sqcup K} F_J G_K$. The sum is over all complementary order-preserving sub-strings $J$ and $K$ of $I$. For example,
    \beq
    (F \circ G)_{ijk} = F_{\emptyset}G_{ijk} + F_i G_{jk} + + F_j G_{ik} + F_k G_{ij} + F_{ij}G_k + F_{ik}G_j + F_{jk} G_i  + F_{ijk}G_\emptyset.
    \eeq
Physically, the shuffle product is the product induced on gluon correlations by the point-wise product of Wilson loop expectation values $\bra (W_1 W_2)(\g) \ket = \bra W_1(\g) \ket \bra W_2(\g) \ket$, when path ordered exponentials are expanded in iterated integrals of gluon correlations in the large-$N$ limit \cite{deform-prod-der}. ${\bf C}\bra \bra A \ket \ket$ with the shuffle product is the shuffle algebra on $\La$ generators $Sh_\La$. The empty word $1$ is a unit element for $sh$, with $1 \circ F = F \circ 1 = F$ for all $F \in Sh_\La$.

Concatenation is defined as $\xi^I \xi^J = \xi^{IJ}$, which extends linearly to $(FG)(\xi) = F_I G_J \xi^{IJ}$. Using the inner product on ${\bf C}\bra \bra A \ket \ket$ for which $\xi^I$ form an orthonormal basis $(\xi^I,\xi^J) = \gd^{I,J}$, we can define the adjoint of concatenation or the deconcatenation coproduct, $\D = conc^\dag$ by $(F,GH) = (\D F, G \otimes H)$. On monomials, $\D \xi^I = \gd^I_{JK} \xi^J \otimes \xi^K$. It is extended linearly to series $\D F = \sum_{J,K} F_{JK} \xi^J \otimes \xi^K$. $\D$ is not co-commutative, it mimics the coproduct on Wilson functions coming from concatenation of loops $(\D W)(\g_1,\g_2) = W(\g_1 \g_2)$. We showed in \cite{deform-prod-der} that $\D$ is a homomorphism of $sh$. The homomorphism of $sh$, $\eps: Sh_\La \to {\bf C}$ which picks out the constant term, $\eps(F) = F_\emptyset$, is a counit. Thus $(sh, conc^\dag = \D, 1, \eps)$ is a bialgebra, the $sh$-$deconc$ bialgebra.

To turn this bialgebra into a Hopf algebra, we define an antipode $S: {\bf C}\bra \bra A \ket \ket \to {\bf C}\bra \bra A \ket \ket$, that mimics the antipode $(SW)(\gamma) = W(\bar \gamma)$ on functions on loop space. On basis elements, let
    \beq
    S(\xi^I) = (-1)^{|I|} \xi^{\bar I}, ~~~ {\rm where~~} \overline{i_1 i_2 \cdots i_n} = i_n \cdots i_2 i_1.
    \eeq
Extend it linearly to $S(F_I \xi^I) = \sum_I (-1)^{|I|} F_{\bar I} \xi^I$. For example, $S(\xi^{i_1 i_2 i_3}) = - \xi^{i_3 i_2 i_1}$.  This comes from comparing the expansions of $W(\gamma)$ and $W(\bar \gamma)$ in terms of gluon correlations. For $S$ to be an antipode it must satisfy several conditions which are usually summarized in a commutative diagram (see \cite{chari-pressley}). (1) It must be a homomorphism of the commutative shuffle algebra. On basis elements, this is the requirement $S(\xi^I \circ \xi^J) = S(\xi^I) \circ S(\xi^J)$
or equivalently,
    \beq
    \sum_{I \sqcup J =K} (-1)^{|K|} \xi^{\bar K} = \sum_{\bar I \sqcup \bar J =L} (-1)^{|L|} \xi^L.
    \eeq
This is indeed true. Riffle shuffling two card packs ($I,J$) preserves the order of each. So reversing the order of the result of the shuffle (each summand on the lhs) is the same as reversing the order of each card pack ($\bar I, \bar J$) and then shuffling them together (each summand on the rhs). The minus signs just come along for the ride; $S$ would be a homomorphism even without them. 

(2) The next two conditions $S(1) =1$ and $\eps S = \eps$ are obviously satisfied. 

(3) The most interesting requirement for $S$ to be an antipode is its compatibility with deconcatenation and shuffle\footnote{If ${\bf C}\bra \bra A \ket \ket$ were the algebra of functions on a group, these conditions would follow from the property that the product of a group element with its inverse in either order is the group identity.}
    \beq
    sh (S \otimes 1) \D &=& sh (1 \otimes S) \D = 1 \eps {\rm ~~~~or ~~equivalently}, \cr
    \gd^I_{JK} (-1)^{|J|} \xi^{\bar J} \circ \xi^{K} &=& \gd^I_{JK} (-1)^{|K|} \xi^J \circ \xi^{\bar K} = \gd^I_{\emptyset} ~~~~ {\rm for~~all~} ~~I.
    \label{e-compatibility-antipode}
    \eeq
Putting $I = i_1 \cdots i_n$ these are the conditions
    \beq
    \sum_{p=0}^n (-1)^p \xi^{i_p \cdots i_1} \circ \xi^{i_{p+1} \cdots i_n} = \sum_{p=0}^n (-1)^{n-p} \xi^{i_1 \cdots i_p} \circ \xi^{i_n \cdots i_{p+1}} = \gd^n_0.
    \eeq
We have not found any nice proof of this, though we verified it explicitly for $n \leq 3$ and observed a pattern of cancelations for higher $n$ which leads us to conjecture that it is an identity. Cartier \cite{cartier-primer} mentions that $sh$-$deconc$ must form a Hopf algebra on general grounds, though we would still like an explicit proof of (\ref{e-compatibility-antipode}). The minus signs in the definition of the antipode are crucial for this compatibility condition to hold. In the sequel we will assume this condition is satisfied.

\section{Matrix model analogue of the group of loops}
\label{s-mat-mod-analogue-of-loop-gp}

The $sh$-$deconc$ Hopf algebra we described is a commutative but non-cocommutative Hopf algebra, so it should be the algebra of functions on some non-abelian group. Which group is it? In the case of Yang-Mills theory, the corresponding group is that of based loops on space time. Remarkably, there seems to be an analogue of this group for hermitian $\La$-matrix models. One might speculate that it is a group built from $U(N)$ or a free group on $\La$ generators (since the concatenation algebra of correlations is the free associative algebra), but this is not the case. Rather, we will construct it as the group of complex valued characters (also known as the dual or spectrum) of the $sh$-$deconc$ Hopf algebra. One might suspect that the analogue of loops are words in the generators of the shuffle algebra; but words do not form a group in a simple-minded way. Nevertheless, we will associate a (family of) group elements to each word and show that they form a subgroup of the spectrum. In another direction, using a result of Ree and Friedrichs \cite{rimhak-ree,friedrichs}, we will identify the spectrum with the exponential of the free Lie algebra.

Consider the set of real/complex-valued characters $\chi$ of the shuffle algebra, which are not identically zero. These are linear homomorphisms from the commutative shuffle algebra to the complex numbers. Suppose $F , G \in Sh$ then
    \beq
    \chi(F \circ G) = \chi(F) \chi(G) {\rm ~~~and~ for~~} a,b \in {\bf C},~~~ \chi(aF + bG)=a\chi(F) + b\chi(G).
    \eeq
It follows that $\chi(1) =1$ for all characters $\chi$. We will define a group structure on this set and call it $\spec(Sh_\La)$ or $\spec_\La$. Suppose $\chi(\xi^I) = \chi^I$. Then the complex numbers $\chi^I$, which we call the character coefficients, completely specify the character. For any $F \in Sh$, $\chi(F) = \chi(F_I \xi^I) = F_I \chi^I$. The $\chi$ form a dual space to the $F \in Sh$, which justifies the upper and lower indices. We can also think of a character as a formal power series $\chi = \chi^I \xi_I$. The identity is taken as the counit $\eps: Sh \to {\bf C}$ defined as $\eps(F) = F_{\emptyset}$, which is a rather trivial homomorphism. In terms of coefficients, $\eps^I = \gd^I_\emptyset$. The product is non-abelian in general and is defined using $\D = conc^\dag$. So it encodes the monoid structure of concatenation. More precisely, $\chi \psi = (\chi \otimes \psi) \D$, which is a map from $Sh \otimes Sh \to {\bf C} \otimes {\bf C}$. Then we identify ${\bf C} \otimes {\bf C}$ with ${\bf C}$ by multiplying the two components, to get a map $Sh \otimes Sh \to {\bf C}$. On basis elements,
    \beq
    (\chi \psi)(\xi^I) = \gd^I_{JK} ~~\chi(\xi^J) ~\psi(\xi^K) = \gd^I_{JK} \chi^J \psi^K.
    \eeq
It is extended linearly to the rest of $Sh$. So the product of characters is just the concatenation product of character power series. The formula for the product does not use the fact that characters are homomorphisms of shuffle. But we need the latter property to show that products of characters are also characters. $\D = conc^\dag$ is a homomorphism of the shuffle product as are $\chi$ and $\psi$. Therefore, $\chi \psi = (\chi \otimes \psi) \D$ is also a homomorphism of the shuffle product: $(\chi \psi)(F \circ G) = (\chi \psi)(F) ~~ (\chi \psi) (G)$.
Indeed, each side is equal to (we identify ${\bf C} \otimes {\bf C}$ with ${\bf C}$)
    \beq
    (\D F)_{I,J} ~(\D G)_{K,L} ~\gd_M^{I \sqcup K} ~\gd_N^{J \sqcup L} ~\chi^M ~\psi^N.
    \eeq
The inverse of a character is defined by composing with the antipode: $\chi^{-1} = \chi S$. $S$ and $\chi$ are homomorphisms of shuffle and so $\chi^{-1}$ is also a homomorphism, and hence a character:
    \beq
    \chi^{-1}(\xi^I) = \chi (S(\xi^I)) = (-1)^{|I|} \chi(\xi^{\bar I}) = (-1)^{|I|} \chi^{\bar I}.
    \eeq
The conditions $\chi \chi^{-1} = \chi^{-1} \chi = \eps$ are precisely the same as the compatibility conditions (\ref{e-compatibility-antipode}) of the antipode $S$ with product $sh$ and coproduct $\D$. Indeed, using the homomorphism property of $\chi$ and the second equality above,
    \beq
    (\chi \chi^{-1})(\xi^I) = \gd^I_{JK} \chi(\xi^J) (-1)^{|K|} \chi(\xi^{\bar K}) = \chi(\gd^I_{JK}
        (-1)^{|K|} \xi^{J} \circ \xi^{\bar K}) = \chi(\gd^I_\emptyset) = \eps(\xi^I).
    \eeq
Similarly we verify that $\chi^{-1} \chi = \eps$. The shuffle algebra is the commutative algebra of functions with pointwise product on the group of characters. The value of a function $F$ at the character $\chi$ is obtained by evaluating the character on $F$:  $F(\chi) \equiv \chi(F) = \chi^I F_I$. Moreover, $(F \circ G)(\chi) = F(\chi) G(\chi)$ since $\chi(F) \chi(G) = \chi(F \circ G)$.

We still need to find non-trivial characters. If $\chi = \chi^I \xi_I$ wants to be a character, $\chi^I$ cannot be arbitrary. On the one hand, $\chi(F) = F_I \chi^I$ may not converge, but we can consider polynomial $F$ so that the series terminates. On the other hand, $\chi$ must be a homomorphism of $sh$, and this imposes relations on the $\chi^I$. For polynomial $F$ and $G$, $\chi(F \circ G) = \chi(F) \chi(G)$ is satisfied iff $\chi(\xi^I \circ \xi^J) = \chi(\xi^I) \chi(\xi^J)$ for all $I,J$ or equivalently
    \beq
         \sum_{I \sqcup J = K} \chi^K = \chi^I \chi^J  ~~~~{\rm for~~ all~~} I,~J.
    \eeq
These conditions were called the shuffle relations in another context \cite{rimhak-ree}. They are the complete set of conditions for $\chi$ to be a character. In detail, the first few shuffle relations are
    \beq
    \chi^\emptyset &=& 1, \cr
    \chi^{ij} + \chi^{ji} &=& \chi^i \chi^j, \cr
    \chi^{ijk} + \chi^{jik} + \chi^{jki} &=& \ch^i \chi^{jk}, \cr
    \chi^{ijkl} + \chi^{ikjl} + \chi^{iklj} + \chi^{kijl} + \chi^{kilj} + \chi^{klij} &=& \chi^{ij} \chi^{kl}, \cr
    \chi^{ijkl} + \chi^{jikl} + \chi^{jkil} + \chi^{jkli} &=& \chi^i \chi^{jkl}, {\rm ~~~ e.t.c.}
    \eeq
For rank $n \geq 1$ character coefficient tensors $\chi^{i_1 \cdots i_n}$, there are $[n/2]$ systems of linear shuffle relations (i.e. either $\half (n-1)$ or $n/2$ according as $n$ is odd or even). The shuffle relations are hierarchical, in the sense that the rank of the tensors on the lhs ($|I|+|J|$) always exceeds the rank of the tensors on the rhs ($|I|$ and $|J|$). So we can think of these as linear equations constraining the higher rank $\chi^K$ in terms of the lower rank ones which appear quadratically as sources on the right. This structure is reminiscent of the matrix model fSDE: $S^{J_1 i J_2} G_{J_1 I J_2} = \gd_I^{I_1 i I_2} G_{I_1} G_{I_2}$ \cite{deform-prod-der,non-anom-ward-id}. Naively, we expect a large space of solutions to these constraints, since there seem to be a lot more degrees of freedom in the $\chi^I$ than there are shuffle relations. In particular, the $\chi^i$ are unconstrained. Regard $\chi^{ij}$ as a matrix. Then its symmetric part is completely determined by the $\chi^i$, but its anti-symmetric part $\half(\chi^{ij} - \chi^{ji})$ is not.

In the case of loop space, we can find examples of characters easily. For example, given a loop $\gamma(t)_{0 \leq t \leq 1}$ on space-time $M$, its value on an element of $F \in Sh(M)$ is $\gamma(F) = \int_\gamma F$ where the rhs is the iterated `Chen' integral\cite{chen-bulletin,tavares} of the linear combination of tensor products of one forms. For example if $F = \a \otimes \gb$ for a pair of $1$-forms $\a$  and $\gb$, then
    \beq
    \gamma(F) = \int_0^1 dt_1~ \int_0^{t_1} dt_2~ \a_i(\g(t_1))~ \gb_j(\g(t_2))~ \dot \g^i(t_1)~ \dot \g^j(t_2).
     \eeq
For the shuffle algebra on a finite number of generators, we might imagine that the analogue of a loop is a word $\xi_I$ and define a linear functional on $Sh$ by $\xi_I(F) = F_I$. However, this is not a character since $\xi_I(1) = \gd^\emptyset_I$, whereas for a character we must have $\chi(1) =1$. Though single words are in the dual of $Sh$ regarded as a vector space, they are not (with the exception of the empty word) in the dual of $Sh$ regarded as an algebra.

Thinking of a character as a formal power series $\chi(\xi) = \sum_I \chi^I \xi_I$, we ask whether there are any characters aside from the identity $\eps$. To begin with, we show using the shuffle relations that there are no non-trivial polynomial characters. Suppose $\chi$ is a non-trivial polynomial character, of degree $n-1$. What this means is that $\chi^K =0$ for all words $K$ of length $|K| \geq n \geq 2$, but with $\chi^I \ne 0$ for some $I$ of rank $|I|=n-1$. Then consider the homomorphism condition $\chi(\xi^I \circ \xi^I) = \chi^I \chi^I$ which is the same as
    \beq
    \sum_{I, I \sqcup I =K} \chi^{K} = \chi^I \chi^I.
    \eeq
The rhs is non-vanishing by assumption. But the lhs vanishes since it is a linear combination of character coefficients of rank $2n-2 \geq n$. Thus we have a contradiction. So the only polynomial character is the identity $\chi = \eps$. To find non-trivial characters, let us specialize first to the case of a single generator.

\subsection{Characters of the shuffle algebra on one generator $Sh_1$}
\label{s-char-of-sh-1}

For $\La =1$, a character is a formal series $\chi = \sum_{n=0}^\infty \chi_n \xi^n$ in one generator $\xi$. The condition that it be a homomorphism of $sh$ is $\chi_0 = 1$, and the following shuffle relations for each $\chi_n, ~n \geq 1$:
    \beq
    {n \choose r} \chi_n &=& \chi_r \chi_{n-r} ~~~ {\rm for~~~} r =1,2,3, \cdots, [n/2].
    \eeq
In more detail, $\chi_0 =1,~~ 2 \chi_2 = \chi_1^2, ~~ 6 \chi_4 = \chi_2^2,~~ 4 \chi_4 = \chi_1 \chi_3,~~ 5 \chi_5 = \chi_1 \chi_4,~~ 10 \chi_5 = \chi_2 \chi_3$ etc. The general solution is a $1$-parameter family $\chi_n = \ov{n!} \chi_1^n$ for $n \geq 0$. We write $\chi = e^{\chi_1 \xi}$. In particular, there are no polynomial characters. Moreover, if $\chi_1 \ne \psi_1$ then $e^{\chi_1 \xi}$ and $e^{\psi_1 \xi}$ are distinct characters as they have different coefficients in their power series expansions. The identity character is got by choosing $\chi_1 =0$, in which case $\chi =1$. The product $\chi \psi$ is the character whose value on monomials is
    \beq
    (\chi \psi)(\xi^n) = \sum_{r=0}^n \chi_{n-r} \psi_r.
    \eeq
$(\chi \psi)_1 = \chi_1 + \psi_1$ so $\chi \psi$ is the character $e^{(\chi_1 + \psi_1) \xi}$, which agrees with the usual rule for multiplying $\chi = e^{\chi_1 \xi}$ and $\psi = e^{\psi_1 \xi}$. The product is abelian, since we have a single generator. The inverse of $\chi = e^{\chi_1 \xi}$ is $\chi^{-1} = e^{- \chi_1 \xi}$. We call the group of characters of $Sh_1$ as $\spec(Sh_1)$ or $\spec_1$ for short. Though space-time has been reduced, in a sense, to a single point, $\spec_1$ is a continuous abelian group parameterized by one real/complex number $\chi_1$. Indeed we can even define a one dimensional abelian Lie algebra on the vector space $\{ \chi_1 \xi ~|~ \chi_1 \in {\bf C} {\rm ~or~} {\bf R} \}$ with Lie bracket $[\chi_1 \xi, \psi_1 \xi] =0$. $\chi = e^{\chi_1 \xi}$ is an exponential map from the Lie algebra to the group.

If we consider real-valued characters, then $f: e^{\chi^1 \xi} \mapsto e^{\chi^1}$ is an isomorphism from $\spec_1$ to the multiplicative group of non-zero reals, ${\bf R}^*$. For complex-valued characters, $f$ is a homomorphism from $\spec_1$ onto ${\bf C}^*$, the multiplicative group of non-zero complex numbers. Its kernel is the subgroup generated by the character $e^{2 \pi i \xi}$, i.e. the subgroup $\{e^{2 \pi i m \xi} ~|~ m \in {\bf Z} \}$.


We think of the shuffle algebra $Sh_1$ as the commutative algebra of functions on the group $\spec_1$. The value of $F = F_n \xi^n$ at $\chi$ is
    \beq
    F(\chi) \equiv \chi(\sum_n F_n \xi^n) = \sum_{n} \chi_n F_n {\rm ~~~or~ equivalently~~} \xi^n(\chi) \equiv \chi(\xi^n) = \chi_n.
    \eeq
The shuffle product is the same as the point-wise product of functions on $\spec_1$, since characters are homomorphisms of the shuffle algebra
    \beq
    (\xi^n \circ \xi^m)(\chi) = \chi(\xi^n \circ \xi^m) = {n+m \choose n} \chi(\xi^{n+m}) = {n+m \choose n} \chi_{n+m} = \chi_n \chi_m
    = \xi^n(\chi) \xi^m(\chi).
    \eeq

\subsection{Group of characters of $Sh_\La$: Pure characters}
\label{s-pure-characters}

We now discuss the group of characters of the $sh$-$deconc$ Hopf algebra on $\La > 1$ generators. Corresponding to the inclusions $\xi_i \hookrightarrow \{ \xi_1, \cdots , \xi_\La \}$ we get $\La$ abelian one-parameter subgroups of $\spec(Sh_\La)$ for free, namely $e^{\chi^1 \xi_1}, ~~e^{\chi^2 \xi_2} ,~~ \cdots ,e^{\chi^\La \xi_\La}$.
For instance, the value of $e^{\chi^3 \xi_3}$ on a basis-element of the shuffle algebra is
    \beq
    (e^{\chi^3 \xi_3})(\xi^I) = \left\{\begin{array}{ll}
                                    {(\chi^3)^n \over n!}  & \hbox{if $I = 333...33$ ($n$ times), and} \\
                                    0 & \hbox{otherwise.}
                                  \end{array}       \right.
    \eeq
We call a character $\chi$ pure if it does not mix the letters $\{\xi_1, \cdots, \xi_\La \}$, i.e. $\chi^I =0$ whenever $\xi^I$ contains at least two distinct letters. The only pure characters are the ones listed above. Another way to look at it is that given a letter $\xi_{i_1}$ from the alphabet $\{\xi_1, \cdots , \xi_\La \}$, we get a $1$-parameter family of pure characters $e^{\chi^{1} \xi_{i_1}}$.

\subsection{Mixed characters}
\label{s-mixed-charac}



A natural question is whether given a word $\xi_I$, it is possible to obtain a (family of) character(s) associated to it? This would in a sense be analogous to associating the character of the shuffle algebra $Sh(M)$, $\int_\gamma F$ to each based loop $\gamma$ on space-time where $F \in Sh(M)$.

We call a character $\chi$ mixed if it is not pure, i.e. if there is a word $\xi_I$ containing at least two distinct letters such that $\chi^I \ne 0$. A large class of mixed characters can be obtained by multiplying pure characters. Given a word $\xi_I= \xi_{i_1} \xi_{i_2} \cdots \xi_{i_n}$ and a sequence of complex numbers $\{ \chi^{1}, \cdots ,\chi^{n} \}$, we define a character via the product of pure characters (products of characters are characters (sec. \ref{s-mat-mod-analogue-of-loop-gp}))
    \beq
    \psi = e^{\chi^{1} \xi_{i_1}}~ e^{\chi^{2} \xi_{i_2}}~ \cdots ~e^{\chi^{n} \xi_{i_n}}.
    \eeq
A word-sequence pair $(\xi_I,\vec{\chi}) = (\xi_{i_1} \cdots \xi_{i_n},\{\chi^1, \cdots \chi^n\})$ determines a character. The inverse of $\chi$ is also of the same form, $\psi^{-1} = e^{-\chi^{1} \xi_{i_n}}~ e^{-\chi^{2} \xi_{i_{n-1}}}~ \cdots e^{\chi^{n} \xi_{i_1}}$ and satisfies $\psi \psi^{-1} = \psi^{-1} \psi = \eps$.


Reduced form of word-sequence pairs: Given a word $\xi_{i_1} \cdots \xi_{i_{j-1}} \xi_{i_j} \xi_{i_{j+1}} \cdots \xi_{i_n}$ and a sequence $(\chi^1, \cdots, \chi^{j-1}, \chi^j, \chi^{j+1}, \cdots, \chi^n)$, we define the reduced form of the pair. If a pair of adjacent letters coincide, $\xi_{i_j} = \xi_{i_{j+1}}$, then delete $\xi_{i_j}$ and $\chi^j$, and replace $\chi^{j+1}$ with $\chi^j + \chi^{j+1}$ to get a new word $\xi_{i_1} \cdots \xi_{i_{j-1}} \xi_{i_{j+1}} \cdots \xi_{i_n}$ and a new sequence $\chi^1, \cdots, \chi^{j-1}, \chi^j + \chi^{j+1}, \cdots, \chi^n$. The resulting character is the same as the original one. Moreover, if any of the numbers $\chi_k$ vanishes, just delete it and the letter $\xi_{i_k}$. Proceeding this way, we get a word whose adjacent letters are always distinct, and a sequence of non-zero complex numbers (the one exception is if the word is empty). Such a word along with its sequence of complex numbers is in reduced form. Of course, the length of the word is the same as the length of the sequence. Two pairs $(\xi_I,\vec{\chi})$ and $(\xi_J,\vec{\psi})$ are equivalent if they have the same reduced forms. Equivalent pairs correspond to the same character. For example, the reduced form of the pair $(\xi_3 \xi_1 \xi_1 \xi_2, \{i,-3,\pi,0 \})$ is $(\xi_3 \xi_1, \{i,\pi-3 \})$ and corresponds to the character $e^{i \xi_3} e^{(\pi-3) \xi_1}$ whose inverse is $e^{(3-\pi) \xi_1} e^{-i \xi_3} $. The reduced form of the identity character is the pair consisting of the empty word and the empty sequence, $(\emptyset,\{\})$. A pure character $e^{\chi^1 \xi_{i_1}}$ has reduced form $(\xi_{i_1},\{\chi^1\})$.

We can multiply two characters corresponding to the words $\xi_{i_1} \cdots \xi_{i_n}$ and $\xi_{i_{n+1}} \cdots \xi_{i_{n+m}}$ to get a character corresponding to the concatenated word $\xi_{i_1} \cdots \xi_{i_{n+m}}$ and the concatenated sequence of complex numbers $\chi^1, \cdots, \chi^{n+m}$:
    \beq
        e^{\chi^{1} \xi_{i_1}}~ e^{\chi^{2} \xi_{i_2}}~ \cdots ~e^{\chi^{n} \xi_{i_n}}~ e^{\chi^{n+1} \xi_{i_{n+1}}}~ e^{\chi^{n+2} \xi_{i_{n+2}}}~ \cdots ~e^{\chi^{n+m} \xi_{i_{n+m}}}.
    \eeq
This product is as non-abelian as it can be. The products of pure characters form a subgroup of the group of characters of $Sh_\La$. From our construction, this subgroup is the free product of $\La$ copies of $\spec_1$
    \beq
    {\rm Group ~of~ prod.~ of ~pure ~charac.} ~~\cong~~ \spec_1 * \spec_1 * \cdots * \spec_1   ~~~~~~ (\La {\rm ~~factors}).
    \eeq
We know that the free group on $n$ generators $F_n$ is the same as the free product of $n$ copies of the integers  $F_1$. By contrast, we will see that $\spec_1 * \spec_1 * \cdots * \spec_1$ is a proper subgroup of $\spec_\La$, at least if we restrict to finite products. It is interesting to know the appropriate topology for such free products of continuous groups ($\spec_1 \cong {\bf R}^*$ for instance).

The exponential of any finite linear combination of generators $\chi = e^{\chi^j \xi_{j}}$ is a character. To show this, consider
    \beq
    \chi = e^{\chi^j \xi_{j}} = \sum_{n=0}^\infty \ov{n!} \chi^{j_1} \xi_{{j_1}} \cdots \chi^{j_n} \xi_{{j_n}}
    = \sum_{n=0}^\infty \ov{n!} \chi^{j_1} \cdots \chi^{j_n} \xi_{{j_1}} \cdots \xi_{{j_n}}.
    \eeq
The coefficients are symmetric tensors $\chi^{j_1 \cdots j_n} = \ov{n!} \chi^{j_1} \cdots \chi^{j_n}$ which must satisfy the shuffle relations $\sum \chi^{I \sqcup J} = \chi^I \chi^J$. Taking $I = i_1 \cdots i_n$ and $J = j_1 \cdots j_m$, on the lhs there are $n+m \choose n$ terms all of which are equal, so
    \beq
    LHS = {n+m \choose n} \chi^{i_1 \cdots i_n j_1 \cdots j_m} = \ov{n! m!} \chi^{i_1} \cdots \chi^{i_n} \chi^{j_1} \cdots \chi^{j_m} = RHS.
    \eeq
$e^{\chi^j \xi_{j}}$ is thus a character, but since $\xi_{j}$ do not commute it cannot be written as a finite product of pure characters. The inverse of $\chi$ is $\chi^{-1} = e^{- \chi^j \xi_{j}}$, which is also of the same form. Since products of characters satisfy the shuffle relations, finite products of the form $\chi = e^{\chi^{i_1} \xi_{i_1}} e^{\chi^{i_2} \xi_{i_2}} \cdots e^{\chi^{i_n} \xi_{i_n}}$ form a group which properly contains $\spec_1 * \cdots * \spec_\La$ and is a proper subgroup of $\spec_\La$. As before, we can put any such product in a reduced form.

Exponentials of arbitrary non-linear polynomials in the generators are not characters in general. For instance, it is easy to see that $\chi = e^{\xi_1 + \xi_{1} \xi_{2}}$ does not satisfy the shuffle relation $\chi^{12} + \chi^{21}= \chi^1 \chi^2$. On the other hand, the Baker-Campbell-Hausdorff formula (BCH) tells us that $e^{\chi^i \xi_i} e^{\psi^j \xi_j} = e^{\chi^i \xi_i + \psi^j \xi_j + \half \chi^i \psi^j [\xi_i,\xi_j] + \ov{12} \chi^i \chi^j \psi^k [\xi_i,[\xi_j,\xi_k]] + \cdots}$. Using BCH we can reexpress products of exponentials occurring in the above subgroup of $\spec_\La$ as exponentials of linear combinations of nested commutators of the generators $\xi_i$. Aside from $\chi^i \xi_i$, these will be certain infinite linear combinations since the $\xi_i$ do not commute. This suggested to us that exponentials of finite (or {\em other} infinite) linear combinations of nested commutators of $\xi_i$ may also be characters. While we verified this in some simple cases, the calculations rapidly get laborious. So we were pleasantly surprised to find this proven in the work of Ree \cite{rimhak-ree} using a theorem of Friedrichs \cite{friedrichs}. More precisely,
linear combinations of iterated commutators of the generators $\xi_i$ are called Lie elements. They are obtained using the operations of taking Lie brackets $[.,.]$ and linear combinations but {\em not} products such as $\xi_i \xi_j$. For example. $C^i \xi_i + C^{ijk} [\xi_i,[\xi_j,\xi_k]] + C^{ijkl} [[\xi_i,\xi_j],[\xi_k,\xi_l]]$ are Lie elements for any tensors $C^{i}, C^{ijk}, C^{ijkl}$. Ree proves that exponentials of Lie elements are the only formal series satisfying the shuffle relations. In other words, the group $\spec_\La$ consists precisely of exponentials of Lie elements. This characterization will be useful since the Schwinger-Dyson operators of an interesting class of matrix models related to Yang-Mills theory, turn out to be Lie elements. Now that we have identified the group $\spec(Sh_\La)$ which plays the role of the group of loops, we can formulate differential calculus on it.

Technically, the group of loops on space-time is a proper subgroup of a larger group of generalized loops (in the language of \cite{tavares}) or extended loop group (in the language of \cite{gambini-pullin-book,bartolo-gambini-griego}). $\spec_\La$ is the analogue of this larger group, while the group generated by pure characters is the analogue of the smaller group of loops. Both in the space-time and matrix model settings, the larger group behaves akin to a classical Lie group and appears to be the correct physical setting for the fSDE. For instance, the SD operators of several matrix models can be interpreted as right-invariant vector fields on the larger group, but not on the smaller one (sec. \ref{s-fSDE-on-grp-spec-La}). Moreover, just as the group generated by pure characters is a free product of $\La$ copies of $\spec_1$, the group of loops is also a free product (free group generated by the based loops).

\section{Differential calculus on the group ${\bf G} = \spec_\La$}
\label{s-calculus-spec-La}

\subsection{Functions}
\label{s-functions}

By a function on the group ${\bf G} = \spec_\La$ we will mean an element of the shuffle algebra $F(\xi) = F_I \xi^I$. Its value at the character $\chi = \chi^I \xi_I \in \spec_\La$ is given by $F(\chi) \equiv \chi(F) = F_I \chi^I$. The ring of such functions is the commutative shuffle algebra $Sh_\La$ with point-wise product of $F$ and $G$ at $\chi$ given by the shuffle product $\sum_I (F \circ G)_I \chi^I$.

\subsection{Vector fields and the Lie algebra of $\spec_\La$}
\label{s-vect-flds-lie-alg-of-spec}

By a vector field $V$ on $\spec_\La$ we will mean a derivation of the shuffle algebra, i.e. a map that takes functions to functions $V : Sh \to Sh$ that is linear over the complex numbers and satisfies the Leibnitz rule $V(F \circ G) = VF \circ G + F \circ VG$. This extends the concept of vector fields to settings more general than differentiable manifolds \cite{madore}. The derivations must form a left-module over the shuffle algebra as well as a Lie algebra over the ring of functions (i.e. $FV$ must be a vector field and the structure functions of the Lie algebra of vector fields must be in $Sh$).

From sec. \ref{s-mixed-charac}, a necessary and sufficient condition for $\chi$ to be in ${\bf G}= \spec_\La$ is that $\log \chi(\xi)$ must be a Lie element, i.e. a linear combination of iterated commutators of $\xi_1, \cdots, \xi_\La$. The set of Lie elements is closed under commutators and forms the free Lie algebra of rank $\La$ (${\rm FLA}_\La$). It is a {\em free} Lie algebra since there are no relations besides linearity, antisymmetry and Jacobi identity that are satisfied by the commutator brackets. This is reminiscent of the exponential map from the Lie algebra to a Lie group. So we expect the Lie algebra of right (or left) invariant derivations of $Sh_\La$ (which should play the role of Lie algebra of the group $\spec_\La$) to be isomorphic to ${\rm FLA}_\La$. We will see that this is indeed the case.

In \cite{deform-prod-der} we showed that linear combinations of iterated commutators of left annihilation $D_i$ are derivations of the shuffle algebra, so they may be considered vector fields on $\spec_\La$. We recall why $D_i$ satisfies the Leibnitz rule: $[D_i (F \circ G)]_I = [F \circ
G]_{iI} = \sum_{I_1 \sqcup I_2 = iI} F_{I_1} G_{I_2}$. Now either $i
\in I_1$ or $i \in I_2$, so
 \beq
    [D_i (F \circ G)]_I &=& \sum_{I_1 \sqcup I_2 = I} (F_{i I_1} G_{I_2}+ F_{I_1} G_{i I_2})
    = \sum_{I_1 \sqcup I_2 = I} ( [D_i F]_{I_1} G_{I_2} + F_{I_1} [D_i G]_{I_2}) \cr
    &=& [(D_i F) \circ G]_I + [F \circ (D_i G)]_I.
 \eeq
Moreover, commutators of left annihilation do not satisfy any relations besides linearity, anti-symmetry and the Jacobi identity. So iterated commutators of $D_i$ span a FLA of rank $\La$. There is a standard basis $D_{(L)}$ for the FLA, labeled by Lyndon words $L$. $D_{(L)}$ is a particular iterated commutator of $D_{l_i}$'s, where $L = l_1 \cdots l_n$ is a Lyndon word (see appendix \ref{a-lyndon-basis-FLA}). So an element of the above FLA of vector fields is written $V = V^L_\emptyset D_{(L)}$ where $V^L_\emptyset$ are (real or complex) constants, and the sum is over all Lyndon words $L$. The value of such a vector field at the point $\chi = \chi^I \xi_I$ is the `tangent vector' $V^L_\emptyset \chi^\emptyset D_{(L)} = V^L_\emptyset D_{(L)}$, since $\chi^\emptyset = 1$ for a character. These `constant coefficient' vector fields can be regarded as forming a sub-algebra of the Lie algebra of $\spec_\La$. For, by evaluating at the identity $\chi = \eps$, they span a space of `tangent vectors' at the identity. We believe these constant coefficient vector fields should be regarded as the whole of the Lie algebra of ${\bf G}$. The structure constants of the FLA of basis vector fields $D_{(L)}$ are denoted $[D_{(L)}, D_{(M)}] = c_{L,M}^N D_{(N)}$, see appendix \ref{a-lyndon-basis-FLA} for examples.

A vector field $V$ on a group is distinguished if it commutes with the action of the group on itself by multiplication (encoded in the coproduct $\D$ on the Hopf algebra of functions). Roughly, this means $\D V$ must equal $V \D$. But that cannot be quite right since $\D : Sh \to Sh \otimes Sh$ while $V: Sh \to Sh$, so we must specify whether $V$ acts on the first or second slot. In fact we must distinguish the right action from the left action, which leads to the definitions $\D V = (V \otimes 1) \D$ and $\D V = (1 \otimes V) \D$ \cite{abe-hopf-algebras}. Alternatively, let $R_g$ be the right translation by $g$ on a group ${\bf G}$ and $R_g^*$ and ${R_g}_*$ the pull-back and push-forward maps. Then the push-forward of a vector field $V$ acts on a function $f$ according to ${R_g}_* V f = {R_g^*}^{-1} V R_g^* f$. Now a vector field is right invariant if ${R_g}_* V = V$ or in other words, $R_g^* V f = V R_g^* f$ for all $f$. But $R_g^*$ is the pull-back induced by right multiplication in the group, and multiplication in the group is encoded in the coproduct $\D$ in the Hopf algebra of functions on ${\bf G}$. This justifies the definitions of right and left invariant derivations by $\D V = (V \otimes 1) \D$ and $\D V = (1 \otimes V) \D$.

If vector fields $V$ and $W$ are right invariant, then so is their product $VW$ (though not a vector field) and hence also their commutator $[V,W]$ (which is a vector field, since commutators of derivations are also derivations). To see this,
    \beq
    (V \otimes 1 ) \D = \D V ~~~\implies~~~ (V \otimes 1 ) \D W = \D V W.
    \eeq
Using the right invariance of $W$ to re-express $\D W$, this becomes
    \beq
    (V \otimes 1 ) (W \otimes 1) \D  = \D V W ~~~\Leftrightarrow~~~ (V W \otimes 1) \D = \D VW.
    \eeq
Thus right invariant derivations form a Lie algebra, which will serve as a substitute for the Lie algebra of right-invariant vector fields on the group ${\bf G}$.

It remains to identify the right-invariant derivations in the case ${\bf G} = \spec(Sh_\La)$. We will show that the constant coefficient vector fields $V = V^L_\emptyset D_{(L)}$ are in fact right invariant. It is straightforward to check that $D_i$ is a right-invariant derivation:
    \beq
    (D_i \otimes 1) \D G &=& (D_i \otimes 1) G_{JK} \xi^J \otimes \xi^K = G_{iJK} \xi^J \otimes \xi^K \cr
    \D (D_i G) &=& \D(G_{iI} \xi^I) = G_{iI} \gd^I_{JK} \xi^J \otimes \xi^K = G_{iJK} \xi^J \otimes \xi^K.
    \eeq
By the previous result, we deduce that iterated commutators of $D_i$ are also right-invariant. For example, $[[D_i,D_j],[D_k,D_l]]$ and $[D_i,[D_j,D_l]]$ are right invariant derivations. Moreover, the condition $\D V = (V \otimes 1) \D$ is linear in $V$, so real/complex linear combinations of iterated commutators of $D_i$ are also right-invariant. In other words, constant coefficient vector fields $V = V^L_\emptyset D_{(L)}$ are right-invariant on ${\bf G}$. In particular, the Schwinger-Dyson operators ${\cal S}^i = 4 g^{ik} g^{jl}[D_j,[D_k,D_l]]$ of Yang-Mills matrix models with action $S = \tr g^{ik} g^{jl} [A_i,A_j][A_k,A_l]$ are right invariant vector fields (see sec. \ref{s-fSDE-on-grp-spec-La}).

A more general derivation of the shuffle algebra is obtained by allowing non-constant coefficients (elements of $Sh$), $V = V^L(\xi) D_{(L)} = V^L_I \xi^I D_{(L)}$. $V$ acts on a function as $VF = V^L(\xi) \circ (D_{(L)} F(\xi))$. The derivation property follows from that of $D_{(L)}$ and commutativity of the shuffle product
    \beq
    V^L(\xi) D_{(L)} (F \circ G) = V^L(\xi) \bigg( D_{(L)} F \circ G + F \circ D_{(L)} G \bigg) = VF \circ G + F \circ VG.
    \eeq
Thus, we can think of $V = V^L(\xi) D_{(L)}$ as a vector field on the group ${\bf G} = \spec_\La$. It is also clear that these derivations of $Sh$ form a left module over the ring of functions on $\spec_\La$. Indeed, shuffle multiplying a derivation $V^L(\xi) D_{(L)}$ from the left by a function $F(\xi)$ gives another derivation $\{F(\xi) \circ V^L(\xi)\} D_{(L)}$.

It is interesting to know whether there are any more derivations of the shuffle algebra; these seem to be adequate for us. We call the space of variable coefficient derivations $Vect({\bf G})$. In a sense, the $D_{(L)}$ form a globally defined moving-frame so that the tangent bundle of ${\bf G}$ is trivial, i.e., $\spec_\La$ is parallelizable just like any Lie group. $Vect({\bf G})$ forms a Lie algebra with Lie bracket (all products are shuffle products)
    \beq
    [V,W] = (V W^L) D_{(L)} - (W V^L) D_{(L)} + c_{L,M}^N V^L \circ W^M D_{(N)}.
    \eeq
Here $V W^L$ is the action of the vector field $V$ on the function $W^L(\xi)$, $V W^L = V^M(\xi) \circ D_{(M)} W^L(\xi)$. We used the fact that $D_{(L)}$ satisfies the Leibnitz rule with respect to $sh$. This is the analogue of the formula for Lie brackets of vector fields on a manifold $[v,w] = v^i \pdr_i w^j \pdr_j - w^i \pdr_i v^j \pdr_j$. $D_{(L)}$ play the role of $\pdr_i$, except they don't commute.


The value of vector field $V = V^L(\xi) D_{(L)}$ at $\chi \in {\bf G}$ is obtained by evaluating the coefficient functions on the character $\chi$, to get a `tangent vector' $V^L_I \chi^I D_{(L)}$. In particular, the value of a vector field at the group identity $\chi^I = \gd^I_\emptyset$ is the `tangent vector' $V^L_\emptyset D_{(L)}$. Thus the space of tangent vectors at the identity is the same as the space of constant coefficient vector fields, which we have also observed to be right-invariant vector fields.  Their Lie algebra is isomorphic to FLA$_\La$; we are now justified to think of it as the Lie algebra of the group ${\bf G} = \spec_\La$.

We argue that non-constant coefficient vector fields $V$ cannot be right-invariant. Suppose there were right-invariant $V$, with $V^L_I \ne 0$ for some non-empty $I$. $V$ as well as the distinct right-invariant vector field with constant coefficients, $V^L_\emptyset D_{(L)}$, both evaluate to the same tangent vector at the identity: $V^L_\emptyset D_{(L)}$. However, there should be a unique right-invariant vector field on ${\bf G}$ that is obtained by right translating the tangent vector $V^L_\emptyset D_{(L)}$, and the constant coefficient vector field $V^L_\emptyset D_{(L)}$ serves that purpose, so $V$ could not be right-invariant. We would still like a combinatorial proof that non-constant coefficient vector fields cannot satisfy $\D V = (V \otimes 1) \D$.

\subsection{One-forms}
\label{s-differential-forms}

A $1$-form is a linear function from $Vect({\bf G})$ to $Sh$. The dual basis of $1$-forms $\tht^{L}$ is also labeled by  Lyndon words $L$. On basis vector fields $\tht^L (D_{(M)}) = \gd^L_M$, and extended linearly to $Vect({\bf G})$, $\tht^L (V^M(\xi) D_{(M)}) = V^L(\xi)$. A general $1$-form $\om = \om_L(\xi) \tht^L$ is a linear combination of the basis $\tht^L$ with coefficients coming from $Sh$. Though $D_{(L)}$ was defined (app. \ref{a-lyndon-words}) through iterated commutators of left annihilation $D_i$, we have not built $\tht^L$ from $\tht^i$. The exterior derivative of an element of the shuffle algebra is a $1$-form defined by its action on vector fields
    \beq
    dF(V^L(\xi) D_{(L)}) = V^L(\xi) \circ D_{(L)} F.
    \eeq
From this we can read off that if $dF = (dF)_L \tht^L$, then the components $(dF)_L = D_{(L)} F$.


\subsection{Differential calculus on $\spec_1$}
\label{s-calculus-spec-1}

Let us illustrate the above formalism in the simplest case of one generator. The spectrum of the shuffle algebra on one generator consists of the exponential series $\chi(\xi) = \sum_{n \geq 0} \chi_n \xi^n$ with $\chi_n = \ov{n!} \chi_1^n$. Functions on $\spec_1$ are elements of the shuffle algebra $F= F_n \xi^n$ with the value $F(\chi) = \sum_{n \geq 0} F_n \chi_n$. $D \xi^n = \xi^{n-1}$ is the only left annihilation operator and its commutators vanish. So the Lyndon basis of the free Lie algebra on one generator is just $D$ and it is an abelian algebra. Moreover $D$ is right invariant since
    \beq
    \D D \xi^n &=& \D \xi^{n-1} = \sum_{p+q=n-1} \xi^p \otimes \xi^q {\rm ~~~~and} \cr
    (D \otimes 1) \D \xi^n &=& (D \otimes 1) \sum_{p+q=n} \xi^p \otimes \xi^q = \sum_{p+q=n} \xi^{p-1} \otimes \xi^q
    \eeq
are equal. The general vector field is $V = V(\xi) D$ where $V(\xi) = \sum_{n \geq 0} V_n \xi^n$ and it is right invariant iff $V(\xi)$ is a constant. $V$ restricts to the tangent vector $\sum_{n \geq 0} V_n \chi_n D$ at the point $\chi$ on the group $\spec_1$ and to the tangent vector $V_0 D$ at the identity.

The 1-form dual to the vector field $D$ is $\tht$ with $\tht(D) =1$. The general $1$-form is $\om = \om(\xi) \tht$. The value of $\om$ on the vector field $V(\xi) D$ is the shuffle product $\om(V) = \om(\xi) \circ V(\xi)$. The exterior derivative of a function is $dF = (DF)~ \tht$ and for a monomial, $d \xi^n = \xi^{n-1} \tht$. The 1-form $\om$ restricts to the co-vector $\sum_{n \geq 0} \om_n \chi_n \tht$ at the point $\chi$ and to the co-vector $\om_0 \tht$ at the identity.

\section{Factorized Schwinger-Dyson equations on ${\bf G} = \spec_\La$}
\label{s-fSDE-on-grp-spec-La}

We found the group $\spec_\La$ that plays the role of the group of based loops, and identified the rudiments of differential calculus on it. Now we'd like to formulate the factorized Schwinger-Dyson equations of matrix models in terms of $\spec_\La$. This may indicate how to generalize the fSDE to groups we are more familiar with, and thereby provide more insight into their solutions. The fSDE ${\cal S}^i G(\xi) = G(\xi) \xi^i G(\xi)$ are a system of equations for the moment generating series $G(\xi) = G_I \xi^I$. $G(\xi)$ is an element of the shuffle algebra or equivalently, a function on the group $\spec_\La$. ${\cal S}^i$ are called the Schwinger-Dyson operators. The only a priori conditions are that $G$ must evaluate to $1$ at the group identity $\chi = \eps$ (i.e. $G_\emptyset = 1$), the coefficients $G_I$ must be cyclically symmetric and satisfy the reality condition $G_I^* = G_{\bar I}$. These follow from the physical requirements of normalization of expectation values $\bra \Ntr 1 \ket =1$, cyclicity of the trace $G_I = \lim_{N \to \infty} \bra \Ntr A_I \ket$ and reality of the matrix model action $S(A)$. It will be useful to keep the gaussian, Chern-Simons and Yang-Mills ($g^{ij} = g^{ji}$) examples in mind,
    \beq
    S_G = \half \tr C^{ij} A_i A_j, ~~~ S_{CS} = {2\sqrt{-1} \kappa \over 3} \tr C^{ijk} A_i[A_j,A_k], ~~~ S_{YM} = g^{ik} g^{jl} [A_i,A_j][A_k,A_l].
    \eeq
Their SD operators were obtained in \cite{deform-prod-der}:
    \beq
    {\cal S}^i_{G} = C^{ij} D_j,~~~~
    {\cal S}^i_{CS} = {\sqrt{-1} \kappa} (C^{ijk} - C^{ikj}) [D_k,D_j],~~~~
    {\cal S}^i_{YM} = 4 g^{ik} g^{jl} [D_j,[D_k,D_l]].
    \label{e-gauss-cs-ym-sd-op}
    \eeq
There is one fSDE for each letter $\xi^i$. But what does a {\em letter} mean in terms of the group? We noticed that letters are primitive elements of the commutative Hopf algebra of functions on ${\bf G}$. Primitive elements $P$ are those that satisfy $\D P = 1 \otimes P + P \otimes 1$, where $\D = conc^\dag$ is deconcatenation, which is {\em not} cocommutative\footnote{In other words $\tau \D \ne \D$, where $\tau(F \otimes G) = G \otimes F$ reverses the order of factors. Note that this is distinct from the Poincare-Birkhoff-Witt construction where the primitives of a {\em cocommutative} Hopf algebra form a Lie algebra whose universal envelope is the Hopf algebra.}. On monomials, $\D \xi^I = \gd^I_{JK} \xi^J \otimes \xi^K$. Let us show that the only primitive elements are linear combinations of letters $\xi^i, 1 \leq i \leq \La$. For an element $P(\xi) = P_I \xi^I$ of the shuffle algebra to be primitive, we need $P_{JK} \xi^J \otimes \xi^K = P_I \xi^I \otimes 1 + 1 \otimes P_I \xi^I$. This is equivalent to the requirements $P_\emptyset =0$ and $\sum_{J,K \ne \emptyset} P_{JK} \xi^J \otimes \xi^K =0$. These conditions are satisfied iff $P_\emptyset = 0$ and $P_I =0$ for $|I| \geq 2$. In other words, linear combinations of letters are the only primitives. So if we pick any basis for the vector space of primitives (such as the letters themselves), we will have one fSDE for each basis element. This is again consistent with the fact that we could rewrite the fSDE as $w_i^j {\cal S}^i G(\xi) = G(\xi) w_i^j \xi^i G(\xi)$ for any non-singular $\La \times \La$ matrix $w^i_j$. This characterization in terms of the primitives of the Hopf algebra of functions on ${\bf G}$ applies to any group.

The rhs of the fSDE $G(\xi) \xi^i G(\xi) = G_H G_J \xi^{HiJ}$ involves concatenation. It can be understood in terms of the convolution product of functions on the group ${\bf G}$. Given a non-abelian group ${\bf G}$, there are two natural dual Hopf algebras associated to it, the commutative algebra ${\bf C}^{\bf G}$ of complex functions $F(g)$ on it with pointwise product $(FG)(g)=F(g) G(g)$, and the non-commutative group algebra ${\bf C} {\bf G}= \{\sum_{g \in {\bf G}} F(g) g \}$ with convolution product $(FG)(g) = \sum_{h \in {\bf G}} F(h) G(h^{-1} g)$. The coproduct of the first becomes the product in the second and vice versa using the duality $\bra \sum_{h \in {\bf G}} F(h)h ,G \ket = \sum_{g \in {\bf G}} F(g) G(g)$. In our case, $sh$-$deconc$ is the commutative Hopf algebra whose coproduct is deconcatenation. The dual Hopf algebra $conc$-$deshuffle$ is the convolution algebra of functions on ${\bf G}$, whose product is concatenation $(FG)_I \xi^I = F_J G_K \xi^{JK}$. Thus $G(\xi) \xi^i G(\xi)$ is the convolution of $G(\xi)$ with the primitive element $\xi^i$ convolved again with $G(\xi)$. This formulation again applies to any group.

For each primitive element $\xi^i$, the lhs of the fSDE is the SD operator ${\cal S}^i$ acting on the function $G(\xi)$ on the group. In the Gaussian, CS and YM examples (\ref{e-gauss-cs-ym-sd-op}), ${\cal S}^i$ is a complex-linear combination of iterated commutators of left annihilation. Let us restrict attention to models where this is the case. Then from our discussion in sec.~\ref{s-vect-flds-lie-alg-of-spec}, we conclude that ${\cal S}^i$ are right invariant vector fields on ${\bf G}$. For this we think of the right-invariant vector fields (or Lie algebra of ${\bf G}$) as represented linearly on the space of functions on the group.

So far, we have formulated practically everything in the fSDE in terms of concepts that generalize to any group without reference to matrix integrals. It only remains to specify which right-invariant vector field ${\cal S}^i$ to associate to a given primitive element $\xi^i$. For this we need additional data beyond the mere specification of a group ${\bf G}$. At present we do not have a prescription of which right invariant vector field ${\cal S}^i$ to associate to a given primitive $\xi^i$, that would apply to an arbitrary group. But we have a rough idea. The additional data is the specification of an action, and the prescription is the passage from action to SD operators. However, even for the group $\spec_\La$ of relevance to matrix models, we only gave $3$ examples (\ref{e-gauss-cs-ym-sd-op}) of actions leading to SD operators which are right-invariant vector fields for each primitive. Moreover, we know that many actions do not lead to SD operators that are right invariant vector fields (eg. $S = \tr A^4$). So we postpone a general characterization of admissible actions and their passage to SD operators, which would apply to any group. Instead, we seek more examples of matrix model actions whose SD operators are right invariant vector fields on $\spec_\La$.

Given a matrix model action $S(A) = \tr S^I A_I$, we define $S(G) = S^J G_J$ with cyclic $S^J$ and $G_J$. The SD operators ${\cal S}^i$ are obtained from the variation in the action under the infinitesimal variations $\gd A_j =L_I^i A_j= \gd_j^i A_I$, i.e., by applying\footnote{$L^i_I$ are infinitesimal automorphisms of the tensor algebra\cite{entropy-var-ppl,deform-prod-der}. Their action on $G_J$ is $L^i_I G_J = \gd^{J_1 i J_2}_J G_{J_1 I J_2}$.} $L_I^i$ to the action\footnote{$|J|$ is the length of the word $J$ and $\bar J$ is the reversed word. Cyclicity of $S^J$ and $G_I$ are used in the fourth equality. $D_j$ is left annihilation, $(D_j G)_I = G_{jI} \implies (D_j D_k G)_I = G_{kjI}$ etc.}
    \beq
    {\cal S}^i G(\xi) &=& \xi^I L_I^i (S^J G_J) = \xi^I S^J \gd_J^{J_1 i J_2} G_{J_1 I J_2} \cr
    &=& \xi^I S^{J_1 i J_2} G_{J_1 I J_2} =  \sum_{J,I} |Ji| S^{iJ} G_{JI} \xi^I
    = \sum_{J,I} |Ji| S^{iJ} (D_{\bar J} G)_I \xi^I.
    \eeq
From this we read off ${\cal S}^i$ (which are not right-invariant vector fields in general)
    \beq
    {\cal S}^i = \sum_J |Ji| S^{iJ} D_{\bar J} = \sum_{n \geq 0} (n+1) S^{i j_1 \cdots j_n} D_{j_n} \cdots D_{j_1}
    \label{e-SD-operator}.
    \eeq
We would like to know which actions lead to ${\cal S}^i$ that are linear combinations of iterated commutators of $D_j$'s, i.e. Lie elements. Linearity of the passage from $S(A)$ to ${\cal S}^i$, implies it is sufficient to work with actions that are homogeneous polynomials of degree $n$ for each $n=2,3,4 \ldots$ separately. One difficulty is that actions are usually presented, for example, in the form  $S(A) = \tr C^{ijk} A_i [A_j,A_k]$ where $C^{ijk}$ are {\em not} cyclic. It takes some relabeling to transform to $S(A) = \tr S^{ijk} A_{ijk}$ with cyclic coupling tensors $S^{ijk}$, in terms of which the SD operators are expressed in (\ref{e-SD-operator}). The other difficulty is to identify those actions for which (\ref{e-SD-operator}) can be rewritten as a linear combination of iterated commutators. In what follows we carry out this program in part and identify a class of actions that lead to ${\cal S}^i$ which are Lie elements.


{\bf Quadratic:} The most general quadratic action is $S_2(A) = \tr C^{ij} A_i A_j$. If we write it as $S_2 = \half \tr (C^{ij} + C^{ji}) A_{ij}$, then the coupling tensor $S_2^{ij} = \half (C^{ij} + C^{ji})$ is cyclically symmetric. By (\ref{e-SD-operator}), ${\cal S}^i = 2 S^{ij} D_j = (C^{ij} + C^{ji}) D_j$. In other words ${\cal S}^i = (C^{ij} + cyclic) D_j$ for all $i$ are Lie elements, i.e. right-invariant vector fields on the group $\spec(Sh_\La)$. Formally, this is true for arbitrary tensors $C^{ij}$, though we should impose the reality condition $S^{ij} = {S^{ji}}^*$ and request that $S^{ij}$ be a positive matrix to ensure that the matrix integrals converge and lead to correlators satisfying $G_I^* = G_{\bar I}$. In the sequel, we will work formally and not mention these reality and positivity conditions.

{\bf Cubic:} Motivated by the Chern-Simons example, we consider the class of cubic matrix models whose action is $S_3(A) = \tr C^{ijk} A_i [A_j,A_k] = \tr C^{ijk}(A_{ijk} - A_{ikj})$ for arbitrary $C^{ijk}$. Writing this in the form $S = \tr (C^{ijk} - C^{ikj}) A_{ijk} \equiv \tr S^{ijk} A_{ijk}$ ensures that the coupling tensor $S^{ijk} = \ov{3} (C^{i[jk]} + {\rm ~cyclic})$ is cyclically symmetric\footnote{The notation $C^{i[jk]} = C^{ijk}-C^{ikj}$.}. We get ${\cal S}^i = 3 S^{ijk} D_k D_j = (C^{ijk} + {\rm ~cyclic}) [D_k,D_j]$ which are Lie elements. Notice that ${\cal S}^i$ is more easily expressed in terms of the original $C^{ijk}$ than in terms of the coupling tensors $S^{ijk}$, which appear as unwelcome middlemen.

{\bf Quartic:} What is the appropriate generalization to higher degree polynomial actions, such that ${\cal S}^i$ remains a Lie element? In \cite{deform-prod-der} we showed that the Yang-Mills type of quartic action $S_{YM} = \tr g^{ik} g^{jl} [A_i,A_j][A_k,A_l]$ leads to the SD operators ${\cal S}_{\rm YM}^i = 4 g^{ik} g^{jl} [D_j,[D_k,D_l]]$ which are Lie elements. However, explicit calculation indicates that the more general $S(A) = \tr B^{ijkl} [A_i,A_j][A_k,A_l]$ leads to SD operators which are {\em not} Lie elements for some $B^{ijkl}$. On the other hand, a quartic generalization of the CS action is $S_4 = \tr C^{ijkl} A_i[A_j,[A_k,A_l]]$. Furthermore, the YM action is a special case of this: using cyclicity of the trace,
    \beq
    \tr g^{ik} g^{jl} [A_i,A_j][A_k,A_l] = \tr g^{ik} g^{jl} (A_{ij[kl]} - A_{i[kl]j}) = \tr g^{ik} g^{jl} A_i[A_j,[A_k,A_l]].
    \eeq
Thus $S_4$ reduces to $S_{YM}$ if $C^{ijkl} = g^{ik} g^{jl}$. This motivates us to check whether the SD operators corresponding to $S_4$ are Lie elements for arbitrary $C^{ijkl}$. Write $S_4$ as
    \beq
    S_4(A) = \tr C^{ijkl}  (A_{ijkl} - A_{ijlk} - A_{iklj} + A_{ilkj}) = \tr (C^{ij[kl]} - C^{il[jk]}) A_{ijkl}
    \eeq
and define $\tilde S^{ijkl} = C^{ij[kl]} - C^{il[jk]}$. Then $S_4 = \tr S^{ijkl} A_{ijkl}$ where the coupling tensor $S^{ijkl} = \ov{4}(\tilde S^{ijkl} + {\rm ~cyclic})$ is cyclically symmetric. Using (\ref{e-SD-operator}) we read off the SD operators ${\cal S}^i = (\tilde S^{ijkl} + {\rm cyclic}) D_{lkj}$. After a lot of relabeling and simplification they can be written as Lie elements ${\cal S}^i = (C^{ijkl} + {\rm ~ cyclic}) [[D_l,D_k],D_j]$! Thus in a sense, $S_4 = \tr C^{ijkl} A_i[A_j,[A_k,A_l]]$ is the proper generalization of $S_{YM}$ while preserving the property that ${\cal S}^i$ be right-invariant vector fields.

{\bf Quintic:} We begin to see a pattern to a class of actions that lead to SD operators that are Lie elements. $S_5 = \tr C^{ijklm} A_i[A_j,[A_k,[A_l,A_m]]]$ is the obvious quintic candidate. After some relabeling, we find
    \beq
    S_5(A) = \tr \tilde S^{ijklm} A_{ijklm}, ~~~ {\rm where~~~} \tilde S^{ijklm} = C^{ijk[lm]} - C^{ijm[kl]} - C^{imj[kl]} + C^{iml[jk]}
    \eeq
so that the cyclic coupling tensor is $S^{ijklm} = \ov{5} (\tilde S^{ijklm} + {\rm ~cyclic})$ and $S_5 = \tr S^{ijklm} A_{ijklm}$. ${\cal S}^i_5 = (\tilde S^{ijklm} + {\rm ~ cyclic}) D_{mlkj}$ after some simplification become
    \beq
    {\cal S}^i_5 = (C^{ijklm} + {\rm ~cyclic}) [[[D_m,D_l],D_k],D_j],
    \eeq
which are Lie elements. While the relation of $C^{ijklm}$ to cyclic coupling tensors $S^{ijklm}$ is non-trivial, the SD operators are simply expressed in terms of $C^{ijklm}$.

{\bf Sixth degree:} For the sixth degree action $S_6(A) = \tr C^{ijklmn} A_i[A_j,[A_k,[A_l,[A_m,A_n]]]]$, the cyclic coupling tensor turns out to be $S^{ijklmn} = \ov{6}(\tl S^{ijklmn} + cyclic)$ where
    \beq
    \tl S^{ijklmn} &=& C^{ijkl[mn]} - C^{ijkn[lm]} - C^{ijnk[lm]} - C^{injk[lm]} \cr && + C^{ijnm[kl]} + C^{injm[kl]} + C^{inmj[kl]} - C^{inml[jk]}.
    \eeq
The corresponding ${\cal S}^i = (C^{ijklmn} + cyclic) [[[[D_n,D_m],D_l],D_k],D_j]$ are again Lie elements.

{\bf Conjecture:} Based on these examples, we conjecture that the $n^{\rm th}$ degree polynomial action
    \beq
    S_n(A) = \tr C^{i_1 \cdots i_n} A_{i_1}[A_{i_2},[A_{i_3},[\cdots[A_{i_{n-1}},A_{i_n}] \cdots ]]]
    \eeq
has SD operators that are the Lie elements
    \beq
    {\cal S}^{i_1}_n = (C^{i_1 \cdots i_n} + cyclic) [[ \cdots [[[D_{i_n},D_{i_{n-1}}],D_{i_{n-2}} ] , D_{i_{n-3}}], \cdots ], D_{i_2}].
    \eeq
We have exhibited a large class of matrix model actions (generalizing the gaussian, Chern-Simons and Yang-Mills ones) whose SD operators are right invariant vector fields on ${\bf G}$. But it may not be an exhaustive list. We have not found the most general Lie elements ${\cal S}^i$ that arise as SD operators of {\em some} action, nor the class of all such actions. Finally, this process must be generalized to other groups; we hope to return to these questions later.


\section{Schwinger-Dyson operators of Yang-Mills theory}
\label{s-sd-op-ym-theory}

Now we illustrate the above framework with the example of Yang-Mills theory. Instead of $\La$ matrices $A_i$, we now have the gluon field, one matrix $A_\mu(x)$ for each space-time point $x$ and $\mu=1, \cdots, d$ where $d$ is the space-time dimension. The sources $\xi^i$ are replaced by $\xi^\mu(x)$. We obtain the factorized Schwinger-Dyson equations in terms of gluon correlations and write the SD operators ${\cal S}^\mu(x)$ (\ref{e-sd-op-ym-theory}) as linear combinations of iterated commutators of left annihilation $D_\mu(x)$ with constant coefficients\footnote{Constant coefficients must be independent of the sources $\xi^\mu(x)$, but could depend on $x$ and $\mu$ which now play the role of the indices $i,j,k$ of matrix models. Coefficients will be differential operators.}. It follows (from sec. \ref{s-vect-flds-lie-alg-of-spec}) that the SD operators of Yang-Mills theory are right-invariant derivations of the shuffle-deconcatenation Hopf algebra generated by the sources $\xi^\mu(x)$ where $x$ and $\mu$ run over all space-time points and polarization indices. However, our formulation is far from complete. The fSDE obtained here must be supplemented by gauge-fixing and ghost contributions for a proper treatment of gauge invariance, before we can look for physical solutions. In \cite{ghost-gluon} we have indicated how to incorporate gauge fixing and ghost terms in the context of matrix models. Mathematically, this means the shuffle-deconcatenation Hopf algebra generated by $\xi^\mu(x)$ needs to be modified. This could be done either by adding generators corresponding to ghosts or by passing to the quotient by an ideal as in Chen's work \cite{chen-bulletin} (see also Tavares \cite{tavares}). Only then can we recover the group of (generalized) loops via the spectrum of the Hopf algebra. We hope to return to these issues in future work, but restrict ourselves here to the pure Yang-Mills action
    \beq
    S = \tr \int d^4x \bigg\{\half \pdr_\mu A_\nu(\pdr^\mu A^\nu ~-~ \pdr^\nu A^\mu)
    ~-~ ig \pdr_\mu A_\nu [A^\mu,A^\nu] ~-~ {g^2 \over 4} [A_\mu,A_\nu][A^\mu,A^\nu] \bigg\}.
    \label{e-ym-action}
    \eeq
To get the SD equations, make the change of integration variable
    \beq
    A_\mu(x) \to A'_\mu(x) = A_\mu(x) + \int v_\mu^{\mu_1 \cdots \mu_n}(x; x_1 \cdots x_n) A_{\mu_1}(x_1) \cdots A_{\mu_n}(x_n) dx_1 \cdots dx_n
    \eeq
in the Euclidean functional integral $Z = \int dA e^{-NS}$.  $v$ are infinitesimal tensors and we work to linear order in them. This is not a local change of variable, but it is not disallowed by any law of physics. If $\gd S$ is the change in action, the Schwinger-Dyson equations relate it to the change in the measure in the sense of expectation values
    \beq
    \bra {\gd S \over N} \ket = \bra \ov{N^2} \bigg\{ \det\bigg({\pdr A' \over \pdr A} \bigg) -1 \bigg\} \ket.
    \eeq
As in sec. \ref{s-intro}, in the limit $N \to \infty$, the factorized SD equations can be written as ${\cal S}^\mu(x) G(\xi) = G(\xi) \xi^\mu(x) G(\xi)$. The generating series of gluon correlations is an element of the shuffle algebra generated by $\xi^\mu(x)$ ($[dx] = dx_1 \cdots dx_n$ where $dx_i$ is the volume element of space-time)
    \beq
    G(\xi) &=&  \sum_{\mu_1, \cdots, \mu_n} \int [dx] G_{\mu_1 \cdots \mu_n}(x_1, \cdots, x_n) \xi^{\mu_1}(x_1) \cdots \xi^{\mu_n}(x_n) \cr
    {\rm where~~~} G_{\mu_1 \cdots \mu_n}(x_1, \cdots, x_n) &=& \lim_{N \to \infty} \bra \Ntr A_{\mu_1}(x_1) \cdots A_{\mu_n}(x_n) \ket.
    \eeq
For the quadratic term on the rhs of the fSDE, we need the jacobian\footnote{Hatted variables (e.g. $\hat{dy_k}$) are not integrated.} $J = \bra \det{(\dd{A'}{A})} \ket$
    \beq
    J&=& 1 + N^2 \int dx [dx] dy_1 \cdots \hat{dy_k} \cdots dy_n
        v_\mu^{\mu_1 \cdots \mu_n}(x;x_1 \cdots x_n) \gd^{\nu_1 \cdots \nu_{k-1} \mu \nu_{k+1} \cdots \nu_n}_{\mu_1 \cdots \mu_n} \cr &&
    \d(x_1 - y_1) \cdots \d(x_{k-1} - y_{k-1}) \d(x_k -x) \d(x_{k+1} - y_{k+1}) \cdots \d(x_{n} - y_{n}) \cr &&
    \bra \Ntr A_{\nu_1}(y_1) \cdots A_{\nu_{k-1}}(y_{k-1}) \ket  \bra \Ntr A_{\nu_{k+1}}(y_{k+1}) \cdots A_{\nu_{n}}(y_{n}) \ket. \cr
    { J -1 \over N^2 }&=&
        \int [dx] \sum_{k=1}^n v_{\mu_k}^{\mu_1 \cdots \mu_n}(x_k;x_1 \cdots x_n) G_{\mu_1 \cdots \mu_{k-1}}(x_1 \cdots x_{k-1}) G_{\mu_{k+1} \cdots \mu_{n}}(x_{k+1} \cdots x_{n}).
    \eeq
The infinitesimal change in action $\bra {\gd S \over N} \ket$ is also linear in the arbitrary tensors $v$. Equating the coefficients of common tensors $v$ leads to the fSDE ${\cal S}^\mu(x) G(\xi) = G(\xi) \xi^\mu(x) G(\xi)$ where the product on the rhs is concatenation. Let us define left annihilation $D_\mu(x)$ (distinct from the covariant derivative) by its action on correlations:
    \beq
    (D_\mu(x) G)_{\mu_1 \cdots \mu_n}(x_1, \cdots, x_n) = G_{\mu \mu_1 \cdots \mu_n}(x,x_1, \cdots, x_n).
    \eeq
Then the Schwinger-Dyson operators are (square brackets denote anti-symmetrization)
    \beq
    {\cal S}^\mu(x) = \pdr_\nu \pdr^{[\mu} D^{\nu]}
        + ig \{ \pdr_\nu [D^\mu,D^\nu] + [\pdr^{[\nu} D^{\mu]}, D_\nu] \}
        - g^2 [D^\nu, [D^\mu,D_\nu]].
    \label{e-sd-op-ym-theory}
    \eeq
For example, let us show how ${\cal S}^\mu(x)$ for the $3$-gluon term $S_3 = - ig \tr \int dx \pdr_\mu A_\nu [A^\mu,A^\nu]$ is obtained. The change in the cubic term is ($[dx]$ stands for $dx dx_1 \cdots dx_n$)
    \beq
    \gd S &=& -ig \tr \int [dx] \pdr_\mu(v_\mu^{\mu_1 \cdots \mu_n}(x;x_1 \cdots x_n) A_{\mu_1}(x_1) \cdots A_{\mu_n}(x_n)) [A^\mu(x),A^\nu(x)] \cr
    && -ig \tr \int [dx] (\pdr_\mu A_\nu(x)) [v^\mu_{\mu_1 \cdots \mu_n}(x;x_1 \cdots x_n) A^{\mu_1}(x_1) \cdots A^{\mu_n}(x_n),A^\nu(x)]  \cr
    && -ig \tr \int [dx] (\pdr_\mu A_\nu(x)) [A^\mu(x),v^\nu_{\mu_1 \cdots \mu_n}(x;x_1 \cdots x_n) A^{\mu_1}(x_1) \cdots A^{\mu_n}(x_n)].
    \eeq
Integrating by parts in the first term we isolate the same $v$ factor in all terms
    \beq
    \gd S &=& ig \tr \int[dx] v^\nu_{\mu_1 \cdots \mu_n}(x;x_1 \cdots x_n) A^{\mu_1}(x_1) \cdots A^{\mu_n}(x_n) \pdr_\mu[A^\mu(x),A_\nu(x)] \cr &&
    -ig \tr \int[dx] v^\nu_{\mu_1 \cdots \mu_n}(x;x_1 \cdots x_n) (\pdr_\nu A_\mu(x)) [A^{\mu_1}(x_1) \cdots A^{\mu_n}(x_n),A^\mu(x)] \cr &&
    -ig \tr \int [dx] v^\nu_{\mu_1 \cdots \mu_n}(x;x_1 \cdots x_n) (\pdr_\mu A_\nu(x)) [A^\mu(x),A^{\mu_1}(x_1) \cdots A^{\mu_n}(x_n)].
    \eeq
We are interested in $\bra {\gd S \over N} \ket$. Before taking expectation values, we should pull $\pdr_x$ and other coupling tensors to the left, while adding additional variables to ensure that derivatives act only on the appropriate fields. This ensures everything can be expressed in terms of gluon correlations
    \beq
    \gd S &=& ig g_{\rho \nu} \tr \int [dx] v^\nu_{\mu_1 \cdots \mu_n}(x;x_1 \cdots x_n) \pdr_\mu^x \bigg\{A^{\mu_1}(x_1)
        \cdots A^{\mu_n}(x_n) [A^\mu(x),A^\rho(x)] \bigg\} \cr &&
    -ig g_{\rho \mu} \tr \int [dxdy] v^\nu_{\mu_1 \cdots \mu_n}(x;x_1 \cdots x_n) \d(x-y) \pdr_\nu^x \bigg\{
        A^\rho(x) [A^{\mu_1}(x_1) \cdots A^{\mu_n}(x_n), A^\mu(y)] \bigg\} \cr &&
    -ig g_{\rho \nu} \tr \int [dxdy] v^\nu_{\mu_1 \cdots \mu_n}(x;x_1 \cdots x_n) \d(x-y) \pdr_\mu^x \bigg\{
        A^\rho(x) [A^\mu(y), A^{\mu_1}(x_1) \cdots A^{\mu_n}(x_n)] \bigg\}
    \eeq
We take expectation values and use cyclicity to move $\mu_1 \cdots \mu_n$ to the right. This facilitates reexpression in terms of left annihilation. Moreover, we omit the common factor $v$ and the integration over $x, x_1 \cdots x_n$, since this facilitates identifying the SD operators
    \beq
    \bra {\gd S \over N} \ket &\propto& ig g_{\rho \nu} \pdr_\mu^x G^{[\mu \rho] \mu_1 \cdots \mu_n} ([x,x],x_1 \cdots x_n)
    + ig g_{\rho \mu} \int dy \d(x-y) \pdr_\nu^x G^{[\rho \mu] \mu_1 \cdots \mu_n} ([x,y],x_1 \cdots x_n) \cr &&
    + ig g_{\rho \nu}  \int dy \d(x-y) \pdr_\mu^x G^{[\mu \rho] \mu_1 \cdots \mu_n} ([y,x],x_1 \cdots x_n) \cr
    &=& ig g_{\rho \nu} \pdr_\mu^x ~([D^\rho,D^\mu]G)^{\mu_1 \cdots \mu_n}(x_1 \cdots x_n) \cr &&
    + ig g_{\rho \mu} \int dy \d(x-y) \pdr_\nu^x ~([D^\mu(y),D^\rho(x)]G)^{\mu_1 \cdots \mu_n}(x_1 \cdots x_n) \cr &&
    + ig g_{\rho \nu} \int dy \d(x-y) \pdr_\mu^x ~ ([D^\rho(x),D^\mu(y)]G)^{\mu_1 \cdots \mu_n}(x_1 \cdots x_n).
    \eeq
We read off the SD operators of the $3$-gluon terms in the Yang-Mills action
    \beq
    {\cal S}_{3}^\mu(x) = ig \bigg\{ \pdr_\nu [D^\mu,D^\nu] + [\pdr^\nu D^\mu - \pdr^\mu D^\nu, D_\nu] \bigg\} =
    ig \bigg\{ \pdr_\nu [D^\mu,D^\nu] + [\pdr^{[\nu} D^{\mu]}, D_\nu] \bigg\}.
    \eeq
Similarly, the SD operators of the quadratic and $4$ gluon terms in the Yang-Mills action (\ref{e-ym-action}),
    \beq
    S_2 &=& \half \tr \int d^4x \pdr_\mu A_\nu(\pdr^\mu A^\nu ~-~ \pdr^\nu A^\mu),  ~~~
    S_4 = -{g^2 \over 4} \tr \int d^4x  [A_\mu,A_\nu][A^\mu,A^\nu]  \cr &{\rm are}& ~~
    {\cal S}^{\mu}_{2}(x) = \pdr_\nu (\pdr^\mu D^\nu - \pdr^\nu D^\mu ) \equiv \pdr_\nu \pdr^{[\mu} D^{\nu]}  \cr &{\rm and}& ~~
    {\cal S}_{4}^\mu(x) = -g^2 g^{\mu \rho} g^{\nu \sigma} [D_\nu , [D_\rho, D_\sigma]] = -g^2 [D^\nu, [D^\mu,D_\nu]].
    \eeq
$S_4$ contains no derivatives that need to be treated with care when identifying coupling tensors. So ${\cal S}_{4}^\mu(x)$ can be read off from the commutator-squared Yang-Mills matrix model of eqn. (\ref{e-gauss-cs-ym-sd-op}). This completes the proof that the SD operators of YM theory are right invariant derivations of the $sh$-$deconc$ Hopf algebra generated by $\xi^\mu(x)$. This result is true independent of space-time dimension.



\section*{\normalsize Acknowledgements}

We would like to thank the EPSRC, UK for support in the form of an EPSRC fellowship.

\appendix

\section{Lyndon words and right standard factorization}
\label{a-lyndon-words}

The SD operators of many interesting matrix models are elements of the free Lie algebra. We identified the FLA with the Lie algebra of right invariant vector fields on the group ${\bf G} = \spec_\La$. To work with these vector fields it is useful to have a basis. Lyndon words introduced by Lyndon \cite{lyndon,chen-fox-lyndon,comb-grp-th,lothaire} are interesting since they label a basis for the FLA. There are other bases for the FLA, such as the Hall basis \cite{hall}. To be self-contained, we summarize some facts (without proofs) about the Lyndon word basis for the FLA.

Suppose we are given the alphabet $\xi^1 , \cdots , \xi^\La$ with the order $\xi^1 < \xi^2 < \cdots < \xi^\La$. There does not seem to be any physically preferred choice for an ordering of the letters. We extend the order on letters to the alphabetical or lexicographic order on all words in the alphabet. For example $\xi^1 \xi^2 < \xi^2 \xi^1$ and $\xi^2 \xi^1 \xi^1 < \xi^2 \xi^1 \xi^1 \xi^3$. If $\xi^I < \xi^J$ we say\footnote{If $I = i_1 i_2 \cdots i_n$ we abbreviate $\xi^{i_1} \xi^{i_2} \cdots \xi^{i_n}$ as $\xi^I$. For brevity we will sometimes talk of the word $L$ when we mean the word $\xi^L$. For instance $I < J$ really means $\xi^I < \xi^J$.} $\xi^I$ precedes $\xi^J$. A Lyndon word is one which is strictly minimal among its conjugates. Conjugates are words related by cyclic permutations; the conjugates of the Lyndon word $\xi^1 \xi^2 \xi^3$ are $\xi^2 \xi^3 \xi^1$ and $\xi^3 \xi^1 \xi^2$. It follows that Lyndon words $\xi^L$ must be primitive, i.e. cannot be written as $(\xi^M)^n$ for some word $\xi^M$ and $n \geq 2$. In particular, a Lyndon word must be aperiodic, since otherwise it would equal one of its non-trivial conjugates. Equivalently, a word is Lyndon iff it precedes every non-empty proper right factor. That is, $\xi^L$ is Lyndon iff for any factorization $\xi^L=\xi^M \xi^N$ with $\xi^M$ and $\xi^N$ non-empty, we have $\xi^L < \xi^N$. Letters are automatically Lyndon words. Lyndon words of length two are $\xi^i \xi^j$ with $\xi^i < \xi^j$. There is also a recursive characterization of Lyndon words: $\xi^L$ is Lyndon iff there exist Lyndon words $\xi^M$ and $\xi^N$ such that $\xi^M < \xi^N$ and $\xi^L = \xi^M \xi^N$. Of course, there may be more than one choice of $M,N$ that do the job. The number of Lyndon words of length $n$ over an alphabet of cardinality $\La$ is given by Witt's formula
    \beq
    l(n,\La) &=& \ov{n} \sum_{d | n} \mu(d) \La^{n/d}, {\rm ~~~ where ~~the~~ Mobius ~~ function} \cr
    \mu(d) &=& \left\{
               \begin{array}{ll}
                 0, & \hbox{if $d$ has a repeated prime factor;} \\
                 1, & \hbox{if $d =1$;} \\
                 (-1)^k, & \hbox{if $d$ is a product of $k$ distinct primes.}
               \end{array}
             \right.
    \eeq
For example, the numbers of Lyndon words of lengths $1$,$2$ and $3$ are given by $l(1,\La) = \La, l(2,\La) = \half (\La^2 - \La), ~~ l(3,\La) = \ov{3}(\La^3 - \La)$.

The {\em right standard factorization} of a Lyndon word $L, |L|>1$ is the unique factorization  $L=MN$ where $M$ and $N$ are Lyndon words such that $N$ is of maximal length. In particular, $M$ must be of length at least one. It follows that $M < L = MN < N$. For clarity we will denote the right standard factorization by $L = M \cdot N$.
For example, $\xi^1 \xi^2 = \xi^1 \cdot \xi^2$ and $\xi^1 \xi^3 \xi^3 = \xi^1 \xi^3 \cdot \xi^3$.

\subsection{Lyndon word basis for free Lie algebra}
\label{a-lyndon-basis-FLA}

We will specify a basis labeled by Lyndon words, for the free Lie algebra (FLA) generated by left annihilation $D_i, 1 \leq i \leq \La$. The FLA consists of linear combinations of iterated commutators of $D_i$. Here we are implicitly thinking of the FLA as embedded in the free associative algebra generated by $D_i$. Elements of the FLA are called Lie elements or Lie polynomials. A Lie element has a definite degree $d$ if it is a homogeneous polynomial of degree $d$ in the free associative algebra. Finding a basis for the FLA is complicated because anti-symmetry and the Jacobi identity relate many different Lie elements. This problem was solved \cite{chen-fox-lyndon,reutenauer} using the right standard factorization of Lyndon words. The basis elements will be called $D_{(L)}$ where $\xi^L$ runs over all Lyndon words. $D_{(L)}$ is a certain iterated commutator of $D_{i}$'s where $\xi^i$ are the letters of the Lyndon word $\xi^L$. We use the notation $D_{(L)}$ to distinguish this iterated commutator from the word $D_L = D_{l_1 \cdots l_n} = D_{l_1} \cdots D_{l_n}$ which is {\em not} a Lie element for $n > 1$.

Letters $\xi^i$ are Lyndon words, and the corresponding degree-one basis elements are $D_i$, which are independent by definition. Given a Lyndon word $L$ with right standard factorization $L = M \cdot N$, we associate the recursively defined Lie element, $D_{(L)} = [D_{(M)},D_{(N)}]$ to the Lyndon word $L$. To express $D_{(L)}$ as an iterated commutator of left annihilation operators, we need to apply this rule recursively till $M$ and $N$ are both single letters. The degree of the basis element $D_{(L)}$ is equal to the length $|L|$. $D_{(L)}$ is not defined when $L$ is not a Lyndon word, though it is sometimes convenient to define $D_{(L)} =0$ if $L$ is not Lyndon.

The Lyndon words of length two are $\xi^i \xi^j, i < j$. Using the right standard factorization $\xi^i \xi^j = \xi^i \cdot \xi^j$ for $i < j$, we get the Lyndon basis $D_{(ij)} = [D_i,D_j]$ for $i < j$. There are clearly $\half (\La^2 - \La)$ of these basis elements. The restriction $i < j$ is explained by the antisymmetry of the commutator.

For an alphabet of two letters ($\La =2$), there are two Lyndon words of length three. Their right standard factorizations are
    \beq
    \xi^1 \xi^1 \xi^2 = \xi^1 \cdot \xi^1 \xi^2  &{\rm with}& \xi^1 < \xi^1 \xi^1 \xi^2 < \xi^1 \xi^2, \cr
    {\rm and ~~~~} \xi^1 \xi^2 \xi^2 = \xi^1 \xi^2 \cdot \xi^2 &{\rm with}& \xi^1 \xi^2 < \xi^1 \xi^2 \xi^2 < \xi^2.
    \eeq
The corresponding Lyndon basis elements are $D_{(112)} = [D_1,[D_1,D_2]]$ and $D_{(122)}= [[D_1,D_2],D_2]$. We see that after accounting for anti-symmetry and the Jacobi identity, there are only two independent Lie elements of degree three for a two letter alphabet. For an alphabet of length $\La =3$, there are $8$ Lyndon words of length three. Six of them involve only two of the letters each and can be obtained from the previous example. We list the $8$ Lyndon basis elements
    \beq
    [D_1,[D_1,D_2]]; ~~~~ [[D_1,D_2],D_2]; && [D_1,[D_1,D_3]]; ~~~~
        [[D_1,D_3],D_3]; \cr
    [D_2,[D_2,D_3]]; ~~~~ [[D_2,D_3],D_3]; && [D_1,[D_2,D_3]]; ~~~~ [[D_1,D_3],D_2].
    \eeq
The right standard factorization of the corresponding Lyndon words may be read off, for example $\xi^{132} = \xi^{13} \cdot \xi^{2}$ and $\xi^{123} = \xi^{1} \cdot \xi^{23}$.

The structure constants of the FLA in the Lyndon basis are defined as $[D_{(L)},D_{(M)}] = c_{L,M}^N D_{(N)}$, where $L,M,N$ are Lyndon words and $|L|+|M|=|N|$. We know of no simple general formula for $c_{L,M}^N$, but give some examples.
    \beq
    [D_i,D_j] &=& \left\{
                  \begin{array}{ll}
                    D_{(ij)}, & \hbox{if $i < j$;} \\
                    D_{(ji)}, & \hbox{if $j < i$;} \\
                    0, & \hbox{if $i=j$.}
                  \end{array}
                \right. \cr
    {\rm Thus ~~for ~~~} |K|=2,~~~
    c_{i,j}^K &=& \left\{
                  \begin{array}{ll}
                    1, & \hbox{if $K=ij$ with $ i < j$;} \\
                    -1, & \hbox{if $K = ji$ with $ j < i$;} \\
                    0 & \hbox{otherwise.}
                  \end{array}
                \right.
    \eeq
Another example is (for $j < k$ so that $jk$ is Lyndon)
    \beq
    [D_i,D_{(jk)}] = \left\{
                     \begin{array}{ll}
               D_{(ijk)}, & \hbox{if $i < jk$;} \\
               D_{(jki)} - D_{(kji)}, & \hbox{if $jk < i$ and $k<i$ and $k<i$;} \\
               D_{(jki)} + D_{(jik)}, & \hbox{if $jk<i$ and $k<i$ and $ji<k$;} \\
               -D_{(jki)}, & \hbox{if $jk<i$ and $i \leq k$.}
                       \end{array}
                     \right.
    \eeq
Another simple example: $[D_{(M)}, D_{(N)}] = D_{(MN)}$ if $MN = M \cdot N$ is the right standard factorization of the Lyndon word $MN$. For this to be the case, it is necessary that $M < MN < N$.



\end{document}